\documentclass[twocolumn]{aastex701}

\usepackage{amsmath}
\begin{document}

\title{A JWST/NIRSpec Catalog of Low-Mass Candidates toward NGC~2024 from Multi-component Spectral Modeling}
 
\correspondingauthor{Shu Wang}
\email{shuwang@nao.cas.cn}

\author[0009-0004-3407-0848]{Haomiao Huang}
\affiliation{CAS Key Laboratory of Optical Astronomy, National Astronomical Observatories, Chinese Academy of Sciences, Beijing 100101, People's Republic of China}
\affiliation{School of Astronomy and Space Sciences, University of Chinese Academy of Sciences, Beijing 100049, People's Republic of China}
\email{huanghm@bao.ac.cn}  

\author[0000-0003-4489-9794]{Shu Wang}
\affiliation{CAS Key Laboratory of Optical Astronomy, National Astronomical Observatories, Chinese Academy of Sciences, Beijing 100101, People's Republic of China}
\affiliation{School of Astronomy and Space Sciences, University of Chinese Academy of Sciences, Beijing 100049, People's Republic of China}
\email[]{shuwang@nao.cas.cn}

\author[0000-0001-7084-0484]{Xiaodian Chen}
\affiliation{CAS Key Laboratory of Optical Astronomy, National Astronomical Observatories, Chinese Academy of Sciences, Beijing 100101, People's Republic of China}
\affiliation{School of Astronomy and Space Sciences, University of Chinese Academy of Sciences, Beijing 100049, People's Republic of China}
\affiliation{Institute for Frontiers in Astronomy and Astrophysics, Beijing Normal University, Beijing 102206, People's Republic of China}
\email{chenxiaodian@nao.cas.cn}

\author[0009-0000-7976-7383]{Zhijun Tu}
\affiliation{CAS Key Laboratory of Optical Astronomy, National Astronomical Observatories, Chinese Academy of Sciences, Beijing 100101, People's Republic of China}
\email{zjtu@bao.ac.cn}

\author{Jifeng Liu}
\affiliation{CAS Key Laboratory of Optical Astronomy, National Astronomical Observatories, Chinese Academy of Sciences, Beijing 100101, People's Republic of China}
\affiliation{School of Astronomy and Space Sciences, University of Chinese Academy of Sciences, Beijing 100049, People's Republic of China}
\affiliation{Institute for Frontiers in Astronomy and Astrophysics, Beijing Normal University, Beijing 102206, People's Republic of China}
\affiliation{New Cornerstone Science Laboratory, National Astronomical Observatories, Chinese Academy of Sciences, Beijing 100012, People's Republic of China}
\email{jfliu@nao.cas.cn}

\begin{abstract}
Young low-mass objects in crowded, highly obscured star-forming regions are difficult to characterize. Deriving reliable physical parameters requires separating photospheric emission from complex extinction and circumstellar emission. We apply a multi-component spectral-modeling framework to JWST/NIRSpec PRISM spectra toward NGC~2024, obtaining adopted fits for 78 sources. The framework incorporates atmospheric models, dust and ice opacities, and optional thermal emission. Most spectra favor additional large-H$_2$O-ice opacity over the baseline small-grain model, while a smaller subset requires additional gray opacity, potentially arising from large refractory grains. Some spectra also require a thermal component to reproduce the long-wavelength continuum. The parameters inferred with this framework, together with the NGC~2024 distance, yield radii and bolometric luminosities. Comparison with evolutionary models for young low-mass and substellar objects provides mass estimates and identifies 46 candidate members, {27} sources classified as background, and {five} uncertain sources. Candidate members show lower fitted surface gravities than likely background stars. Of the candidate members, {23} lie above the youngest available isochrones, plausibly reflecting the cluster's youth and uncertainties in model initial conditions. The resulting catalog provides a uniform characterization of the NGC~2024 sample and establishes a framework for treating complex extinction and circumstellar emission in JWST spectra of other obscured star-forming regions.

\end{abstract}


\section{Introduction}
\label{sec:introduction}

Gravitational collapse, fragmentation, and subsequent accretion in molecular clouds produce objects spanning a broad mass range from stars and brown dwarfs to free-floating planetary-mass objects. The origin of the lowest-mass objects remains uncertain. They may form through the same processes as stars, but disk fragmentation, dynamical ejection, and external feedback may also be important \citep[{and references therein}]{2012ARA&A..50...65L}. These pathways may produce different early accretion histories, internal structures, and evolutionary states. Studying newly formed low-mass and substellar objects is therefore essential for understanding how star formation extends across the stellar--planetary mass boundary.

Young embedded clusters provide access to these objects at early stages of their evolution. However, unrelated field stars complicate membership identification, and strong, spatially variable extinction limits optical characterization of the faintest members. Near-infrared spectroscopy of low-mass young objects partly overcomes this limitation because temperature-sensitive molecular features remain observable at high extinction \citep{2007ApJ...657..511A}. These spectra provide the basis for estimating effective temperature, extinction, and bolometric luminosity, which connect the observed sources with pre-main-sequence and substellar evolutionary models \citep{2014ApJ...786...97H}.

The observed near-infrared spectrum is not determined by the photosphere alone. Line-of-sight extinction depends on the grain properties, and how the infrared extinction law varies with environment, especially in dense clouds, still requires observational calibration \citep{2019ApJ...877..116W,2024ApJ...964L...3W}: grain growth can change the extinction curve \citep{1984ApJ...285...89D,2015ApJ...811...38W}, while ice mantles add broad absorption bands \citep{2015ARA&A..53..541B}. Circumstellar material may also add thermal emission, altering the continuum and weakening photospheric features. These effects are coupled and can produce degeneracies among the fitted parameters. A model consisting only of a photosphere and a standard extinction law may therefore bias the inferred physical parameters. Reliable parameter inference therefore requires a multi-component framework that simultaneously models the atmospheric spectrum, dust and ice opacity, and circumstellar thermal emission.

NGC~2024 is a natural test case for such multi-component modeling. It is a very young embedded cluster in Orion~B at a distance of approximately \(415\) pc \citep{1982AJ.....87.1213A,2017ApJ...834..142K}. The cluster exhibits high and spatially variable extinction \citep{2000AJ....120.1396H,2003ApJ...598..375S}, and also hosts a substantial circumstellar-disk population \citep{2020A&A...640A..27V}. Previous near-infrared spectroscopy identified low-mass stars and brown dwarfs down to approximately \(0.02\,M_\odot\) \citep{2006ApJ...646.1215L}, while recent JWST imaging has identified candidates extending into the planetary-mass regime \citep{2025ApJ...981L..34D}. Its rich low-mass population, strong extinction, circumstellar emission, and projected background sources make NGC~2024 well suited for evaluating whether reliable physical parameters can be recovered in a complex star-forming environment.

Here we analyze archival JWST/NIRSpec Micro-Shutter Assembly (MSA) spectra \citep{2022A&A...661A..80J,2022A&A...661A..81F} spanning \(0.6\)--\(5.3~\mu{\rm m}\) for sources projected toward NGC~2024. This broad wavelength coverage allows our multi-component framework to combine BT-Settl model atmospheres \citep{2011ASPC..448...91A,2012RSPTA.370.2765A}, separate dust and ice opacities, and optional thermal emission. We use progressively complex model families, adding large-ice or large-dust opacity only when required by the spectra. From the fitted effective temperature, extinction, and flux scaling, we derive radius and bolometric luminosity and estimate masses using pre-main-sequence and substellar evolutionary models \citep{2000A&A...358..593S,2015A&A...577A..42B,2020A&A...637A..38P}. The same framework can be applied to JWST spectroscopic samples in other obscured star-forming regions. Section~2 describes the data and source sample. In Section~3, we present the spectral modeling framework. Section~4 presents the derived physical properties and source classifications, and Section~5 discusses the interpretation of the modeling results. We summarize our conclusions in Section~6.

\section{Data and Sample}
\label{sec:data}

We analyze pipeline-produced \texttt{x1d} spectra from archival JWST/NIRSpec MSA observations of sources projected toward NGC~2024 (GO program 5409; PI: M. De Furio). {The program was designed to spectroscopically confirm brown-dwarf and planetary-mass candidates identified from prior JWST/NIRCam imaging \citep{2025ApJ...981L..34D}. These proposed candidates had color-based mass estimates below \(40\,M_{\rm Jup}\)\footnote{\url{https://www.stsci.edu/jwst/phase2-public/5409.pdf}}, and the spectroscopy was intended to establish cluster membership and refine these estimates.}  The observations used the PRISM/CLEAR configuration, providing continuous coverage over approximately \(0.6\)--\(5.3~\mu{\rm m}\) at a wavelength-dependent resolving power of \(R\sim30\)--300. 

The \texttt{x1d} products were processed with JWST calibration pipeline version 1.20.2 {\citep{2025zndo..17515973B}} and CRDS context \texttt{jwst\_1464.pmap}. We independently processed the lower-level data with the official JWST calibration pipeline and obtained spectra consistent with the archival \texttt{x1d} products. We therefore adopt the archival products for the present analysis. {The pipeline subtracts a combined background estimated from the other two nod positions within the same MSA configuration before spectral extraction. Residuals from spatially varying nebular emission remain in some spectra.} Source-specific masking and quality control are applied during the spectral fitting described in Section~\ref{subsec:spectral_fitting}. Of the 137 extracted spectra, 85 are retained for forward modeling after excluding those with incomplete wavelength coverage or severe data-quality problems. The spectra of sources included in both this work and \citet{Luhman2026} generally agree. The few sources showing larger differences have poor-quality archival spectra and had already been excluded during our initial selection; none are included in the 85 source modeling sample.

\section{Spectral Modeling and Model Selection}
\label{sec:modeling}

\subsection{Spectral Model}
\label{subsec:spectral_model}

We fit each spectrum with a multi-component forward-modeling framework based on previous Bayesian analyses of JWST/NIRSpec brown-dwarf spectra \citep{2025ApJS..281...49T}. The model in this study jointly combines a BT-Settl photosphere, wavelength-dependent dust and ice opacity, and an optional phenomenological thermal component representing warm circumstellar emission. The model flux is
\begin{equation}
\begin{aligned}
F_{\nu,{\rm model}}(\lambda)
={}&
\left[
10^{\log(R^2/D^2)}
F_{\nu,{\rm phot}}^{\rm surf}(\lambda)
+F_{\nu,{\rm th}}(\lambda)
\right]\\
&\times 10^{-0.4A_\lambda}.
\end{aligned}
\end{equation}
where \(F_{\nu,{\rm phot}}^{\rm surf}\) is the emergent surface flux of the photospheric model. We use solar-composition (\([{\rm M/H}]=0\), \([\alpha/{\rm Fe}]=0\)) BT-Settl spectra \citep{2011ASPC..448...91A,2012RSPTA.370.2765A}, {evaluated} over \(T_{\rm eff}=500\)--\(10000\) K and \(\log g=3.0\)--\(6.0\). {Within the native grid, we interpolate linearly in flux.} For \(T_{\rm eff}<2000\) K, the native grid does not extend below \(\log g=3.5\); we therefore extrapolate each wavelength point to \(\log g=3.0\) linearly in logarithmic flux using the two lowest-gravity grid spectra, {to preserve positive fluxes in the extrapolated spectra}. The BT-Settl spectra are converted from their native \(F_\lambda\) representation to \(F_\nu\) before comparison with the observed spectra. The flux-scaling parameter, \(\log(R^2/D^2)\), is fitted simultaneously with the atmospheric and extinction parameters. {We choose BT-Settl for its broad coverage in effective temperature and surface gravity, including low-gravity atmospheres relevant to young objects. This enables a uniform analysis of the stellar and substellar sources in our sample. In addition, previous JWST/NIRSpec analyses of late-M/L dwarfs yield similar effective temperatures across different atmospheric models \citep{2025ApJ...980..230T}.}

The optional thermal component \(F_{\nu,\rm th}\) combines multi-temperature blackbody emission, using a prescription similar to the DuCKLinG framework \citep{2024A&A...687A.209K}. Its intrinsic spectrum is 
\begin{equation}
F_{\nu,{\rm th}}(\lambda)
=
10^{\log C_{\rm th}}
\int_{T_{\rm out}}^{T_{\rm in}}
B_\nu(\lambda,T)
\left(\frac{T}{T_{\rm in}}\right)^{-2/q_{\rm th}-1}
\,dT,
\end{equation}
where \(B_\nu\) is the Planck function. The radial temperature profile follows \(T(r)\propto r^{-q_{\rm th}}\). {Representative passive-disk models give \(q_{\rm th}=3/7\) for flared disks and \(3/4\) for flat disks \citep{1997ApJ...490..368C}.} The coefficient \(C_{\rm th}\) scales the integral to the observed flux density. Models without this component adopt \(F_{\nu,{\rm th}}=0\). This phenomenological prescription describes possible disk-related continuum excess without specifying a unique disk structure.

The wavelength-dependent extinction is parameterized as
\begin{equation}
A_\lambda =
A_{K,\rm small}\,\kappa_{\rm small}(\lambda)
+
A_{K,\rm large}\,\kappa_{\rm large}(\lambda),
\end{equation}
where both opacity curves, \(\kappa_{\rm small}(\lambda)\) and \(\kappa_{\rm large}(\lambda)\), are normalized to unity at \(2.2~\mu{\rm m}\). Thus \(A_{K,\rm small}\) and \(A_{K,\rm large}\) describe the \(K\)-band extinction contributions from the baseline small-grain component and the additional large-opacity component, respectively. The fitted \(A_{K,\rm small}\) and \(A_{K,\rm large}\), together with their corresponding extinction curves, are used to derive \(A_{V,1}\) and \(A_{V,2}\) at \(0.55~\mu{\rm m}\).

The baseline component contains graphite and silicate \citep{1984ApJ...285...89D,2003ApJ...598.1017D}, H$_2$O-rich mixed ice, and pure CO ice. We adopt the standard approximation of homogeneous spherical grains \citep{1984ApJ...285...89D}, treating each species as a separate population. For each population, the wavelength-dependent extinction cross sections are calculated from the adopted optical constants using Mie theory. The extinction cross sections of the individual components are then combined. Scattering is therefore treated as removal from the direct beam rather than through a full radiative-transfer calculation. 

{The first term of Equation~3, corresponding to the small-grain extinction (\(A_{K,\rm small}\kappa_{\rm small}(\lambda)\)), is modeled using} {a power-law} size distribution, \(dn/da\propto a^{-3.5}\) \citep{1977ApJ...217..425M}, extending from a fixed \(a_{\min}= 3.5\) \AA\ to a fitted \(a_{\max,\rm small}\). The lower cutoff \(a_{\min}\) follows the interstellar grain-size distributions of \citet{2001ApJ...548..296W}, which extend to ultrasmall grains. The refractory dust is parameterized by its total solid volume fraction \(f_{\rm dust}\). The solid volume ratio of graphite to silicate is fixed at \(f_{\rm C}:f_{\rm sil}=1:1.5\), consistent with the relative volumes inferred from elemental abundances and depletions by \citet{2001ApJ...548..296W}. This gives \(f_{\rm C}=0.4f_{\rm dust}\) and \(f_{\rm sil}=0.6f_{\rm dust}\). The ice opacity includes an H$_2$O-rich mixed-ice component with H$_2$O:CO$_2$:CO:NH$_3$=100:16:8:8 \citep{2022A&A...666A.153D}. This fixed mixture represents the H$_2$O-dominated molecular environments commonly inferred for interstellar ice mantles \citep{2015ARA&A..53..541B}. We also include a pure-CO component \citep{1997A&A...328..649E}. The solid-volume weights of the mixed-ice and pure-CO components are \(f_{\rm H_2O}\) and \(f_{\rm CO}=1-f_{\rm dust}-f_{\rm H_2O}\), respectively. Combinations with \(f_{\rm dust}+f_{\rm H_2O}>1\) are rejected.

\begin{deluxetable}{llc}
\tabletypesize{\footnotesize}
\tablewidth{0pt}
\tablecaption{{Spectral Model Families}\label{tab:model_families}}
\tablehead{
\colhead{{Family}} &
\colhead{{Components}} &
\colhead{\({k}\)}
}
\startdata
{SG} & {Photosphere + small grains} & {8} \\
{SG+TH} & {SG + thermal emission} & {12} \\
{LI} & {SG + large H$_2$O ice} & {11} \\
{LI+TH} & {LI + thermal emission} & {15} \\
{LD} & {LI + large refractory dust} & {14} \\
\enddata
\tablecomments{{Small grains denote the baseline refractory dust and ice extinction component. TH denotes thermal emission. \(k\) is the total number of free parameters, including \(\log f\).}}
\end{deluxetable}

\begin{deluxetable*}{llll}
\tabletypesize{\footnotesize}
\tablewidth{0pt}
\tablecaption{Standard Priors for the Spectral Models\label{tab:model_priors}}
\tablehead{
\colhead{Parameter} &
\colhead{Description} &
\colhead{Prior} &
\colhead{Applicability}
}
\startdata
\(T_{\rm eff}\) (K) & Effective temperature & \(\mathcal{U}(500,10000)\) & All \\
\(\log g\) & Surface gravity & \(\mathcal{U}(3.0,6.0)\) & All \\
\(\log(R^2/D^2)\) & Photospheric flux scaling & \(\mathcal{U}(-30,-10)\) & All \\
\(A_{K,\rm small}\) (mag) & Small-grain extinction magnitude & \(\mathcal{U}(0,10)\) & All \\
\(f_{\rm dust}\) & Dust volume weight & \(\mathcal{U}(0,1)\) & All \\
\(f_{\rm H_2O}\) & Mixed-ice volume weight & \(\mathcal{U}(0,0.4)\) & All \\
\(a_{\max,\rm small}\) (\(\mu{\rm m}\)) & Maximum small-grain radius & \(\mathcal{U}(0.1,2.0)\) & All \\
\(A_{K,\rm large}\) (mag) & Large-opacity amplitude & \(\mathcal{U}(0,10)\) & LI, LD \\
\(a_{0,\rm H_2O,vlg}\) (\(\mu{\rm m}\)) & Characteristic radius of large ice & \(\mathcal{U}(0.5,10)\) & LI, LD \\
\(\sigma_{\rm H_2O,vlg}\) & Width of large-ice size distribution & \(\mathcal{U}(0.1,1.0)\) & LI, LD \\
\(f_{\rm dust,vlg}\) & Large-dust volume weight & \(\mathcal{U}(0,1)\) & LD \\
\(a_{0,\rm dust,vlg}\) (\(\mu{\rm m}\)) & Characteristic radius of large dust & \(\mathcal{U}(0.5,10)\) & LD \\
\(\sigma_{\rm dust,vlg}\) & Width of large-dust size distribution & \(\mathcal{U}(0.1,1.0)\) & LD \\
\(A_{K,\rm large}f_{\rm dust,vlg}\) & Large-dust contribution proxy & \(p\propto\exp[-A_{K,\rm large}f_{\rm dust,vlg}/0.05]\) & LD \\
\(T_{\rm in}\) (K) & Inner disk temperature & \(\mathcal{U}(500,2000)\) & TH \\
\(T_{\rm out}\) (K) & Outer disk temperature & \(\mathcal{U}(150,T_{\rm in})\) & TH \\
\(q_{\rm th}\) & Disk temperature profile exponent & \(\mathcal{U}(0.1,1.0)\) & TH \\
\(\log C_{\rm th}\) & Disk flux scaling & \(\mathcal{U}(-4,4)\) & TH \\
\(\log f\) & Fractional error inflation & \(\mathcal{U}(-10,0)\) & All \\
\enddata
\tablecomments{
\(\mathcal{U}(a,b)\) denotes a uniform prior between \(a\) and \(b\). In the Applicability column, All indicates parameters used in all five model families. LI denotes parameters used in LI and LI+TH, LD those used only in LD, and TH those shared by the two thermal families, SG+TH and LI+TH.
}
\end{deluxetable*}

The second term, \(A_{K,\rm large}\kappa_{\rm large}(\lambda)\), represents an additional large-opacity component. Large ice and large dust follow lognormal size distributions,
\begin{equation}
\frac{dn}{da}
\propto
\frac{1}{a\sigma}
\exp\left[-\frac{\ln^2(a/a_0)}{2\sigma^2}\right],
\end{equation}
following the very-large graphite and H$_2$O-ice grain prescriptions of \citet{2015ApJ...811...38W} and \citet{2015MNRAS.454..569W}, respectively. Here, \(a_0\) is the characteristic radius and \(\sigma\) is the width in \(\ln a\). The opacity is averaged per unit solid volume over 3.5 \AA\ \(\leq a\leq10~\mu{\rm m}\). Large dust uses the same graphite/silicate mixture as the small-grain dust, while large ice is treated as pure H$_2$O ice using optical constants from the Leiden Ice Database \citep{2022ApJ...925..179H,2022A&A...668A..63R}. We do not use CO$_2$-bearing large ice because Mie calculations using the mixed-ice optical constants of \citet{2022A&A...666A.153D} produce additional scattering structure near the 4.27~\(\mu{\rm m}\) CO$_2$ band that is not seen in the observed spectra. In models with both large dust and large ice, \(f_{\rm H_2O,vlg}=1-f_{\rm dust,vlg}\), {where the subscript ``vlg'' denotes very large grains}. Because the large-ice and large-dust cross sections are mixed before the combined opacity is normalized at \(2.2~\mu{\rm m}\), we use \(A_{K,\rm large}f_{\rm dust,vlg}\) only as a proxy for the strength of the large-dust contribution. This relatively gray contribution is degenerate with parameters such as \(\log(R^2/D^2)\); we therefore apply an exponential prior \(p \propto \exp[-A_{K,\rm large}f_{\rm dust,vlg}/\tau_{\rm vlg}]\), with \(\tau_{\rm vlg}=0.05\). {Because large-dust extinction lacks distinctive near-infrared features and is degenerate with photospheric flux scaling, this prior shifts posterior weight toward weaker large-dust contributions, favoring the smallest contribution consistent with the observed spectrum.}

We compare five model families. The baseline model contains only small-grain opacity and is referred to as the small-grain model (SG). Adding large-ice opacity to SG defines another family, namely the large-ice model (LI). Extending LI with large refractory-dust opacity gives the large-dust model (LD). The corresponding thermal extensions of SG and LI are referred to as the small-grain plus thermal model (SG+TH) and large-ice plus thermal model (LI+TH), respectively. {Table~\ref{tab:model_families} summarizes their components and numbers of free parameters.}

\begin{figure*}
\centering
\includegraphics[width=\textwidth]{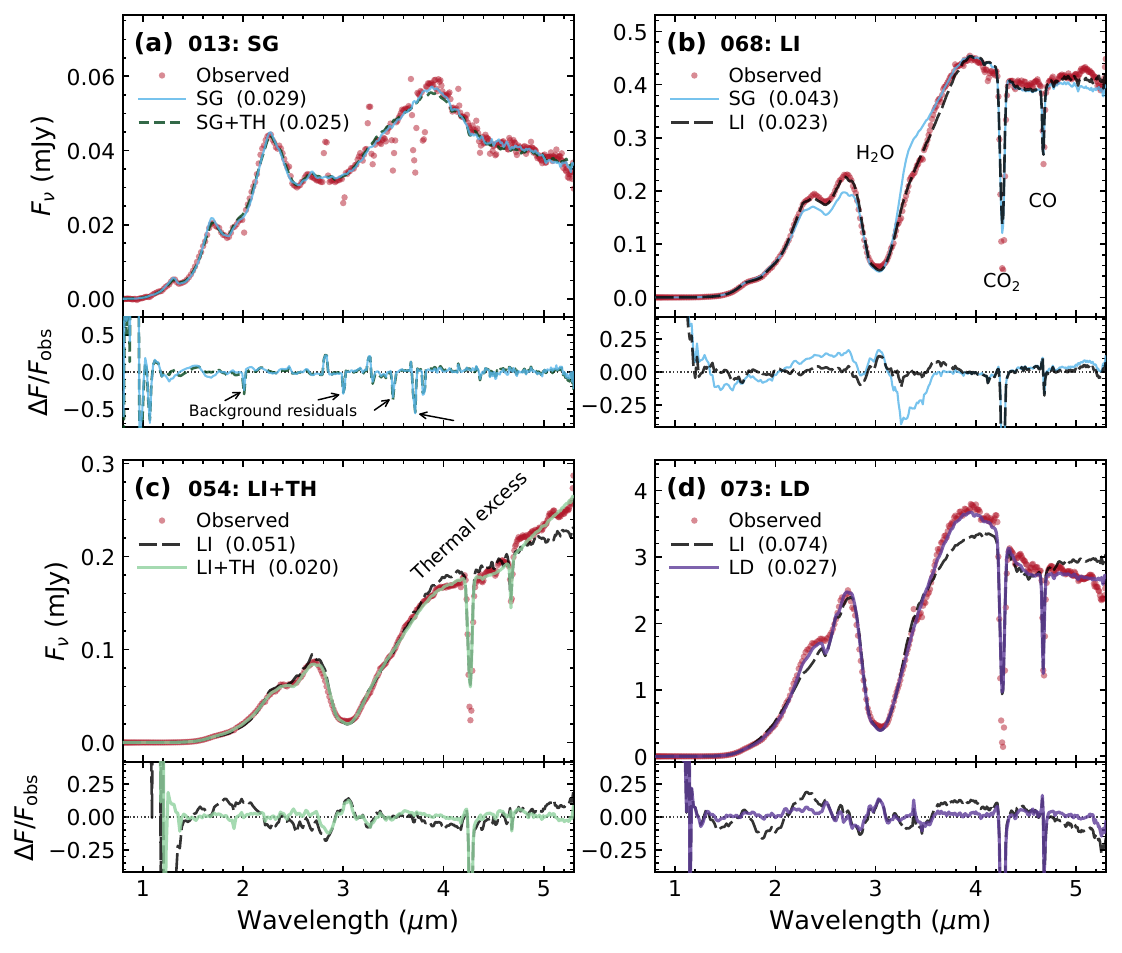}
\caption{
Representative NIRSpec/PRISM spectra illustrating the model hierarchy. Red points show the observed spectra, colored curves show the fitted models, and the lower panels show the fractional residuals, \(\Delta F/F_{\rm obs}\), where \(\Delta F=F_{\rm obs}-F_{\rm model}\). The panel titles give the source ID and adopted model family. The legends show the models being compared, with their MAF values in parentheses. Panels (a)--(d) illustrate the adopted SG, LI, LI+TH, and LD families, respectively. SG denotes the baseline small-grain model. LI consists of SG plus large-H$_2$O-ice opacity. TH denotes the thermal component. LD consists of LI plus large-dust opacity. {Arrows in the lower panel of (a) indicate part of the background-subtraction residuals. Labels mark the H$_2$O, CO$_2$, and CO ice absorption bands in (b) and the long-wavelength thermal excess in (c).}}
\label{fig:model_hierarchy}
\end{figure*}

\subsection{Fitting Procedure and Model Selection}
\label{subsec:spectral_fitting}

\begingroup
\setlength{\tabcolsep}{3pt}
\centerwidetable
\begin{deluxetable*}{lccccccccccccc}
\tablecaption{Adopted Spectral and Physical Parameters\label{tab:source_catalog}}
\tabletypesize{\scriptsize}
\tablewidth{0pt}
\tablehead{
\colhead{ID} & \colhead{Class} & \colhead{Model} & \colhead{$T_{\rm eff}$} & \colhead{$\log g$} & \colhead{$A_{K,\rm small}$} & \colhead{$A_{K,\rm large}$} & \colhead{$f_{\rm th}^{1-2.6}$} & \colhead{$R$} & \colhead{$\log(L_{\rm bol})$} & \colhead{$M_{\rm ATMO}$} & \colhead{$M_{\rm BHAC}$} & \colhead{$M_{\rm Siess}$} & \colhead{${\Delta{\rm BIC}}$} \\
\colhead{} & \colhead{} & \colhead{} & \colhead{K} & \colhead{} & \colhead{mag} & \colhead{mag} & \colhead{\%} & \colhead{$R_\odot$} & \colhead{$L_\odot$} & \colhead{$M_{\rm Jup}$} & \colhead{$M_{\rm Jup}$} & \colhead{$M_{\rm Jup}$} & \colhead{}
}
\startdata
003 & O & LI+TH & $2525_{-35}^{+38}$ & $3.14_{-0.11}^{+0.21}$ & $2.49_{-0.01}^{+0.01}$ & $0.16_{-0.01}^{+0.01}$ & 1.35 & $0.84_{-0.03}^{+0.04}$ & $-1.60_{-0.03}^{+0.03}$ & $13.8_{-1.2}^{+1.4}$ & $16.3_{-1.8}^{+2.4}$ & \nodata & ${230.78}$ \\
004 & B & LI & $4784_{-135}^{+152}$ & $4.94_{-0.52}^{+0.38}$ & $2.08_{-0.02}^{+0.02}$ & $0.14_{-0.01}^{+0.02}$ & 0 & $0.087_{-0.004}^{+0.004}$ & $-2.45_{-0.05}^{+0.05}$ & \nodata & \nodata & \nodata & ${0.98}$ \\
009 & {B$^{*}$} & {SG+TH} & ${2425_{-23}^{+20}}$ & ${3.42_{-0.20}^{+0.15}}$ & ${1.28_{-0.03}^{+0.04}}$ & \nodata & {13.07} & ${0.12_{-0.01}^{+0.01}}$ & ${-3.33_{-0.04}^{+0.04}}$ & {\nodata} & \nodata & \nodata & ${6.30}$ \\
013 & O & SG & $1959_{-5}^{+4}$ & $3.07_{-0.04}^{+0.06}$ & $0.86_{-0.01}^{+0.01}$ & \nodata & 0 & $0.25_{-0.01}^{+0.01}$ & $-3.11_{-0.03}^{+0.03}$ & $5.49_{-0.50}^{+0.50}$ & \nodata & \nodata & ${3.34}$ \\
016 & M & SG+TH & $2375_{-35}^{+34}$ & $3.12_{-0.09}^{+0.18}$ & $2.30_{-0.04}^{+0.04}$ & \nodata & 0.04 & $0.29_{-0.01}^{+0.01}$ & $-2.61_{-0.04}^{+0.04}$ & $10.5_{-0.9}^{+0.9}$ & $11.3_{-0.8}^{+1.5}$ & \nodata & ${2.53}$ \\
017 & B & {LI} & ${5200_{-201}^{+171}}$ & ${4.23_{-0.77}^{+0.73}}$ & ${2.39_{-0.01}^{+0.01}}$ & ${0.06_{-0.01}^{+0.01}}$ & 0 & ${0.077_{-0.003}^{+0.003}}$ & ${-2.42_{-0.06}^{+0.05}}$ & \nodata & \nodata & \nodata & ${23.06}$ \\
054 & M & LI+TH & $3113_{-46}^{+62}$ & $4.31_{-0.27}^{+0.28}$ & $3.03_{-0.04}^{+0.07}$ & $0.24_{-0.02}^{+0.02}$ & 1.97 & $0.40_{-0.02}^{+0.02}$ & $-1.87_{-0.04}^{+0.04}$ & $68.4_{-7.0}^{+6.1}$ & \nodata & \nodata & ${193.54}$ \\
057 & B$^{*}$ & LI+TH & $2977_{-116}^{+116}$ & $3.82_{-0.44}^{+0.49}$ & $1.65_{-0.04}^{+0.05}$ & $0.09_{-0.01}^{+0.01}$ & 56.22 & $0.18_{-0.02}^{+0.02}$ & $-2.65_{-0.10}^{+0.10}$ & \nodata & \nodata & \nodata & ${88.72}$ \\
061 & {M} & {LI} & ${2613_{-12}^{+15}}$ & ${4.50_{-0.06}^{+0.06}}$ & ${1.07_{-0.01}^{+0.01}}$ & ${0.04_{-0.01}^{+0.01}}$ & 0 & ${0.25_{-0.01}^{+0.01}}$ & $-2.59_{-0.03}^{+0.03}$ & ${20.1_{-1.0}^{+0.8}}$ & \nodata & \nodata & ${8.98}$ \\
067 & {U} & {LI+TH} & ${2796_{-211}^{+210}}$ & ${4.79_{-1.02}^{+0.76}}$ & ${1.30_{-0.11}^{+0.10}}$ & ${0.09_{-0.01}^{+0.01}}$ & {92.81} & ${0.24_{-0.05}^{+0.06}}$ & ${-2.49_{-0.32}^{+0.29}}$ & ${31.1_{-12.9}^{+15.8}}$ & \nodata & \nodata & ${41.77}$ \\
068 & O & LI & $2533_{-34}^{+44}$ & $3.09_{-0.07}^{+0.14}$ & $3.04_{-0.04}^{+0.05}$ & $0.21_{-0.01}^{+0.02}$ & 0 & $0.96_{-0.04}^{+0.04}$ & $-1.47_{-0.04}^{+0.04}$ & $14.0_{-1.2}^{+1.6}$ & $16.8_{-1.8}^{+2.9}$ & \nodata & ${6.43}$ \\
069 & B$^{*}$ & LI & $6466_{-173}^{+180}$ & $3.69_{-0.50}^{+0.82}$ & $4.18_{-0.02}^{+0.02}$ & $0.15_{-0.02}^{+0.03}$ & 0 & $0.48_{-0.02}^{+0.02}$ & $-0.45_{-0.05}^{+0.05}$ & \nodata & \nodata & \nodata & ${55.05}$ \\
071 & M & LD & $3313_{-77}^{+134}$ & $3.22_{-0.13}^{+0.17}$ & $3.00_{-0.02}^{+0.02}$ & $1.32_{-0.22}^{+0.20}$ & 0 & $3.32_{-0.52}^{+0.50}$ & $0.07_{-0.09}^{+0.09}$ & \nodata & $245_{-30}^{+63}$ & $293_{-28}^{+55}$ & ${96.47}$ \\
073 & M & LD & $3246_{-90}^{+109}$ & $3.64_{-0.27}^{+0.34}$ & $3.51_{-0.03}^{+0.03}$ & $1.90_{-0.18}^{+0.17}$ & 0 & $4.97_{-0.64}^{+0.67}$ & $0.39_{-0.08}^{+0.08}$ & \nodata & $219_{-35}^{+44}$ & $269_{-29}^{+40}$ & ${152.86}$ \\
074 & O & LI+TH & $2845_{-70}^{+75}$ & $3.39_{-0.25}^{+0.31}$ & $3.18_{-0.05}^{+0.07}$ & $0.22_{-0.02}^{+0.02}$ & 11.87 & $1.10_{-0.08}^{+0.09}$ & $-1.16_{-0.05}^{+0.05}$ & $31.1_{-6.1}^{+9.6}$ & $85.0_{-33.8}^{+21.9}$ & \nodata & ${21.74}$ \\
079 & O & LD & $2429_{-25}^{+25}$ & $3.06_{-0.05}^{+0.11}$ & $2.70_{-0.02}^{+0.02}$ & $0.93_{-0.08}^{+0.10}$ & 0 & $0.91_{-0.06}^{+0.07}$ & $-1.59_{-0.05}^{+0.06}$ & $11.6_{-0.9}^{+0.9}$ & $12.1_{-1.0}^{+1.1}$ & \nodata & ${66.47}$ \\
090\tablenotemark{a} & O & LD & $2300_{-90}^{+94}$ & $3.20_{-0.15}^{+0.27}$ & $2.86_{-0.09}^{+0.09}$ & $0.24_{-0.07}^{+0.12}$ & 0 & $0.25_{-0.02}^{+0.02}$ & $-2.82_{-0.06}^{+0.07}$ & $9.21_{-1.54}^{+3.06}$ & $13.2_{-2.4}^{+2.3}$ & \nodata & ${5.03}$ \\
093\tablenotemark{a} & O & LI+TH & $1521_{-16}^{+34}$ & $5.43_{-0.13}^{+0.05}$ & $9.91_{-0.15}^{+0.07}$ & $1.60_{-0.06}^{+0.06}$ & 76.57 & $2.23_{-0.30}^{+0.27}$ & $-1.62_{-0.12}^{+0.10}$ & $3.19_{-0.32}^{+0.34}$ & \nodata & \nodata & ${138.59}$ \\
101 & M & LI & $2790_{-60}^{+56}$ & $4.49_{-0.21}^{+0.20}$ & $4.67_{-0.05}^{+0.05}$ & $0.17_{-0.02}^{+0.02}$ & 0 & $0.66_{-0.03}^{+0.03}$ & $-1.63_{-0.04}^{+0.04}$ & $26.1_{-4.2}^{+5.0}$ & $48.4_{-10.4}^{+11.2}$ & \nodata & ${22.97}$ \\
140 & B$^{*}$ & LI+TH & $3796_{-243}^{+141}$ & $5.63_{-0.60}^{+0.27}$ & $3.82_{-0.33}^{+0.22}$ & $0.24_{-0.06}^{+0.05}$ & 8.67 & $0.48_{-0.10}^{+0.06}$ & $-1.37_{-0.22}^{+0.10}$ & \nodata & \nodata & \nodata & ${2.13}$ \\
$\cdots$ & $\cdots$ & $\cdots$ & $\cdots$ & $\cdots$ & $\cdots$ & $\cdots$ & $\cdots$ & $\cdots$ & $\cdots$ & $\cdots$ & $\cdots$ & $\cdots$ & $\cdots$ \\
\enddata
\tablenotetext{a}{The spectral fit is flagged; its parameters should be used with caution.}
\tablecomments{The table is published in its entirety in machine-readable format. A portion is shown here for guidance regarding its form and content. Class codes are M (matched), O (overluminous), B (background), B$^{*}$ (background by the evolutionary-model criterion with independent evidence suggesting membership), and U (uncertain). The seven B$^{*}$ sources remain part of the {27}-source Background class. Model codes follow the definitions in {Table~\ref{tab:model_families}}. Except for \(f_{\rm th}^{1-2.6}\), the values are posterior medians with 16th--84th percentile intervals. The reported parameter intervals do not include atmospheric model-grid interpolation uncertainties. The \(f_{\rm th}^{1-2.6}\) column gives maximum-likelihood percentages used for classification; its posterior summaries are included in the machine-readable table. Masses from the three evolutionary models are reported independently, with the posterior and Hayashi-track uncertainties added in quadrature. {The final column gives \(\Delta{\rm BIC}={\rm BIC}_{\rm next}-{\rm BIC}_{\rm adopted}\), where \({\rm BIC}_{\rm next}\) is the lowest BIC among the other model families, excluding the two TH families for Source~069; positive values favor the adopted model. The complete machine-readable table additionally provides mass-estimation methods, fit-quality and wavelength-mask flags, BIC values for all tested model families, and fit statistics for the adopted model.}}
\end{deluxetable*}
\endgroup

For the standard fits, we use the \(0.8\)--\(5.3~\mu{\rm m}\) wavelength interval and include only points with finite fluxes and uncertainties and \({\rm S/N}>10\). Some spectra contain residuals near the \(3.3~\mu{\rm m}\) PAH band and the adjacent \(3.4\)--\(3.6~\mu{\rm m}\) aliphatic hydrocarbon features \citep{2023ApJ...959...74B}. These features and other source-dependent nebular lines are masked individually. Background-subtraction artifacts and visually identified bad pixels are also masked. For a small number of sources, we adjust the fitted wavelength range or temperature prior after inspecting the initial fit. In particular, wavelengths below \(1.5~\mu{\rm m}\) are masked for several highly extincted sources with anomalous short-wavelength continua likely caused by imperfect background subtraction.

The model and extinction curves are evaluated on a fine wavelength grid with \(\Delta\lambda=0.001~\mu{\rm m}\). After applying the extinction, the model is smoothed with Gaussian kernels and resampled onto the observed PRISM wavelength grid.

For a given model family, the fit is evaluated with a Gaussian likelihood. To account for residual calibration errors, slit-loss uncertainties, and {mismatches associated with the stellar-atmosphere and dust/ice extinction models}, we include a fractional error-inflation parameter \(\log f\). The variance at each fitted wavelength point is
\begin{equation}
s_i^2 =
\sigma_{F,i}^2
+
(10^{\log f}M_i)^2
+
\sigma_{\rm floor}^2 ,
\end{equation}
where \(F_i\), \(\sigma_{F,i}\), and \(M_i\) are the observed flux, its uncertainty, and the model flux, respectively. The last term \(\sigma_{\rm floor}=10^{-12}\) Jy prevents unrealistically small variances for very faint fluxes. The log-likelihood is then
\begin{equation}
\ln \mathcal{L}=-\frac{1}{2}\sum_i\left[(F_i-M_i)^2/s_i^2+\ln(2\pi s_i^2)\right].
\end{equation}
The standard parameter priors are listed in Table~\ref{tab:model_priors}; a small number of source-specific refits use restricted temperature priors. Posterior distributions are sampled with UltraNest nested sampling \citep{2021JOSS....6.3001B}. {Batch fits for model comparison use \(\texttt{min\_num\_live\_points}=400\) and \(\texttt{min\_ess}=200\). For the final adopted fits,} we report the posterior median and 16th--84th percentile interval for each fitted parameter. These intervals do not include the uncertainty introduced by linear interpolation within the atmospheric model grid. Tests of linear interpolation in brown-dwarf atmosphere grids indicate that interpolation residuals are generally small over most wavelengths \citep{2024ApJ...976...82T}.

\begin{figure*}
\centering
\includegraphics[width=\textwidth]{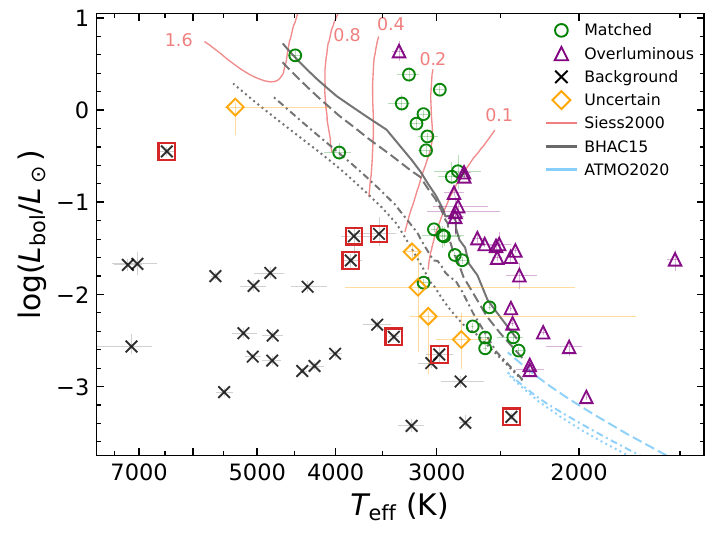}
\caption{
Effective temperature and bolometric luminosity of the 78 fitted sources, initially evaluated at the adopted NGC~2024 distance. Green circles, purple triangles, black crosses, and orange diamonds denote ``Matched'', ``Overluminous'', ``Background'', and ``Uncertain'' sources, respectively. Red open squares identify seven sources classified as ``Background'' despite independent evidence of NGC~2024 membership. Red curves show the pre-main-sequence mass tracks of \citet{2000A&A...358..593S}, labeled in units of \(M_\odot\). Gray and blue curves show the BHAC15 and ATMO~2020 isochrones, respectively \citep{2015A&A...577A..42B,2020A&A...637A..38P}. For clarity, the ATMO~2020 isochrones are shown only below \(2450\) K and the BHAC15 isochrones only above \(2350\) K; the full model grids are used in the mass estimates. Solid, dashed, dash-dotted, and dotted curves correspond to 0.5, 1, 5, and 10~Myr, respectively; the 0.5~Myr curve is available only for BHAC15. Error bars indicate the 16th--84th percentile posterior intervals.
}
\label{fig:teff_lbol}
\end{figure*}

{We first compare the four models, SG, SG+TH, LI, and LI+TH for each source. Degeneracies with the stellar normalization and other extinction components can leave the large-dust contribution poorly constrained in otherwise adequate fits. We therefore consider LD only for sources whose broad spectral shapes are not adequately reproduced by the four models.} The primary residual metric is the median absolute fractional residual (MAF),
\begin{equation}
{\rm MAF} =
{\rm median}_i
\left(
\frac{|F_i-M_i|}{|F_i|}
\right),
\end{equation}
computed over the fitted wavelength points. {MAF summarizes the typical fractional mismatch while limiting the influence of a small number of large residuals.} {We use \({\rm MAF}=0.05\) as a reference threshold for fit quality, together with visual inspection of coherent residuals.} {For relative comparisons among competing models, we use the Bayesian information criterion (BIC; \citealt{1978AnSta...6..461S}),}
\begin{equation}
{{\rm BIC}=k\ln n-2\ln\mathcal{L}_{\max}},
\label{eq:bic}
\end{equation}
{where \(k\) is the number of free parameters and \(n\) is the number of fitted spectral points. All families are compared on the same wavelength points for each source. We estimate \(\ln\mathcal{L}_{\max}\) from the highest Gaussian log-likelihood among the sampling points. For each source, we adopt the model family with the lowest BIC. The exception is Source~069, for which the TH solutions assign nearly all of the flux to thermal emission and yield physically implausible photospheric parameters. We therefore exclude both TH families for this source and adopt LI, the lowest-BIC non-TH model. MAF and visual inspection are used to assess absolute fit quality and coherent residual structure. }

{Of the 85 spectra considered for fitting, seven are excluded after model evaluation, leaving 78 sources with adopted models. Final posterior estimates are obtained from fine fits of the adopted families using \(\texttt{min\_num\_live\_points}=2000\) and \(\texttt{min\_ess}=1600\) to improve posterior sampling and obtain more stable parameter estimates and credible intervals. Eight sources have \({\rm MAF}>0.05\) and are flagged for cautious use.} {Figure~\ref{fig:model_hierarchy} compares models for four representative sources. Source~013 is adequately reproduced by SG, which is favored over SG+TH by BIC. In Sources~068, 054, and 073, the LI, LI+TH, and LD models improve the ice-band profile, long-wavelength continuum, and broad continuum shape, respectively, and are favored by BIC.}

\section{Physical Properties and Source Classification}
\label{sec:results}

The spectral fits require different combinations of dust, ice, and thermal components. The final sample comprises {5 SG, 6 SG+TH, 23 LI, 33 LI+TH}, and 11 LD fits. A thermal component is included in {39} adopted models. Table~\ref{tab:source_catalog} summarizes the adopted model family, fitted parameters, and derived physical properties. Uncertainties are reported as 16th--84th percentile intervals, and the complete table is available in machine-readable form.

\begin{figure}
\centering
\includegraphics[width=\columnwidth]{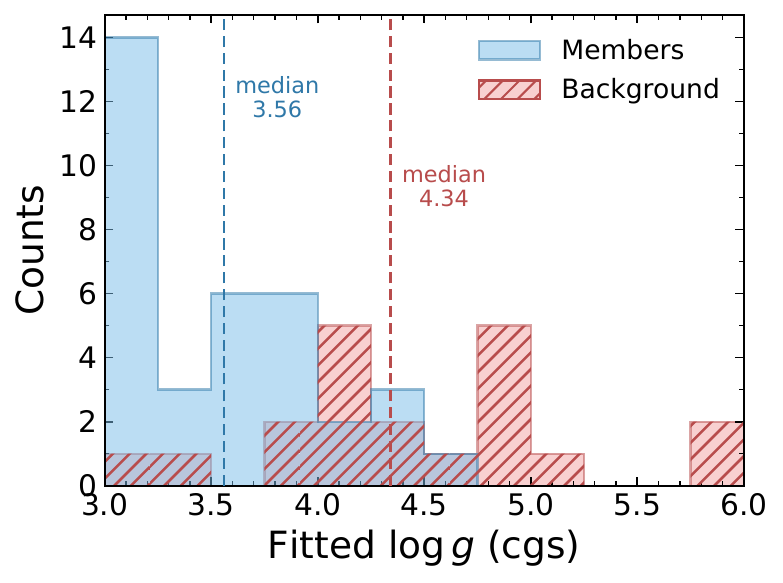}
\caption{
Distributions of the fitted \(\log g\) posterior medians for sources selected using the maximum-likelihood criterion
\(f_{\rm th}^{1-2.6}<0.1\). Blue {unhatched} and red {hatched} histograms represent {35} candidate NGC~2024 members and the {20} ``Background'' sources remaining after the seven B$^{*}$ sources {with independent evidence suggesting NGC~2024 membership} are excluded, respectively. {Both distributions use a bin width of 0.25 dex.} Dashed lines mark the corresponding median values. The fitted \(\log g\) values are used only as a population-level classification check.
}
\label{fig:logg_classification}
\end{figure}

We quantify the short-wavelength thermal contribution using \(f_{\rm th}^{1-2.6}=\int F_{\lambda,\rm th}\,d\lambda/\int(F_{\lambda,\rm phot}+F_{\lambda,\rm th})\,d\lambda\), where both integrals are evaluated over \(1.0\)--\(2.6~\mu{\rm m}\). This fraction measures the contribution of the fitted thermal component to the total model continuum over this wavelength range. The parameters of the thermal component are often weakly constrained and strongly correlated. As a result, a model constructed from their individual marginal posterior medians does not always reproduce the observed continuum. We therefore classify sources using \(f_{\rm th}^{1-2.6}\) from the joint maximum-likelihood solution, which best matches the observed continuum. Using the maximum-likelihood values, sources with \(f_{\rm th}^{1-2.6}<0.1\) are classified as photosphere dominated, those with \(0.1\leq f_{\rm th}^{1-2.6}<0.3\) are flagged as intermediate cases, and those with \(f_{\rm th}^{1-2.6}\geq0.3\) are classified as thermal-excess dominated. {The sample contains 59 photosphere-dominated sources, 5 intermediate cases, and only 14 thermal-excess-dominated sources.} Photospheric features can be weak in thermal-excess-dominated spectra. Estimates of \(T_{\rm eff}\), extinction, flux scaling, and derived physical parameters may therefore be less reliable.

\subsection{Radius and Bolometric Luminosity}
\label{subsec:radius_luminosity}

We initially derive \(R\) and \(L_{\rm bol}\) for every fitted source by placing it at the distance of NGC~2024, enabling a uniform comparison with evolutionary models. Published estimates of this distance include 415 pc from photometry of associated B-type stars \citep{1982AJ.....87.1213A}, \(423\pm15\) pc from VLBA astrometry \citep{2017ApJ...834..142K}, and a recent VLBA estimate of \(403\pm5\) pc \citep{2026A&A...708A.299D}. To encompass the range of published measurements, we adopt a Gaussian distance prior centered at 415~pc with a standard deviation of 15~pc, \(D\sim\mathcal{N}(415,15^2)\)~pc. The radius is derived from \(\log(R^2/D^2)\), with the distance uncertainty propagated through the posterior samples. The bolometric luminosity is calculated as \(L_{\rm bol}=4\pi D^2\int F_{\nu,\rm phot}^{\rm scaled}\,d\nu\), integrating the scaled BT-Settl photospheric spectrum over the full wavelength range of the native model, approximately \(0\)--\(1000~\mu{\rm m}\). The thermal component is excluded from this integration. {For background stars, the radii and luminosities derived at the assumed cluster distance should not be interpreted as their true physical values.} The resulting \(R\) and \(L_{\rm bol}\) values are also reported in Table~\ref{tab:source_catalog}.

\subsection{Evolutionary Models and Mass Estimates}
\label{subsec:evolutionary_masses}

Figure~\ref{fig:teff_lbol} compares the inferred \(T_{\rm eff}\) and \(L_{\rm bol}\) with three sets of evolutionary models: ATMO~2020 \citep{2020A&A...637A..38P}, BHAC15 \citep{2015A&A...577A..42B}, and the pre-main-sequence models of \citet{2000A&A...358..593S}. The grids span complementary mass ranges of \(0.001\)--\(0.075\,M_\odot\), \(0.01\)--\(1.4\,M_\odot\), and \(0.1\)--\(7\,M_\odot\), respectively. Although surface gravity is included in the fit, many substellar models provide only limited constraints on absolute logg for late-M and L dwarfs in low resolution near-infrared spectra \citep{2026arXiv260426771T}. We therefore estimate mass separately with each evolutionary model rather than from \(M=gR^2/G\). The resulting masses differ among the three models and are reported individually without averaging. For posterior samples lying within a model grid, we interpolate in \(T_{\rm eff}\) and \(L_{\rm bol}\) to derive the mass distribution.

Several low-mass sources lie above the youngest available isochrone at their inferred \(T_{\rm eff}\). The estimated age of NGC~2024, approximately \(0.5\) Myr \citep{2006ApJ...646.1215L}, is comparable to the youngest BHAC15 boundary at \(0.5\) Myr and below the \(1\) Myr lower age limit of ATMO~2020; the Siess models extend to a nominal age of \(0\) Myr. Evolutionary predictions at these ages also depend strongly on the assumed initial conditions \citep{2002A&A...382..563B}. Such high luminosities are therefore physically plausible. During early evolution along Hayashi tracks, \(T_{\rm eff}\) generally changes less with age than \(L_{\rm bol}\). For these sources, we derive mass from the relation between \(T_{\rm eff}\) and mass at the youngest model age, provided that \(T_{\rm eff}\) remains within the native model range. This procedure does not extrapolate the model in luminosity. We adopt half the range of masses inferred from the three youngest available model ages as the systematic uncertainty.

\begin{figure*}
\centering
\includegraphics[width=\textwidth]{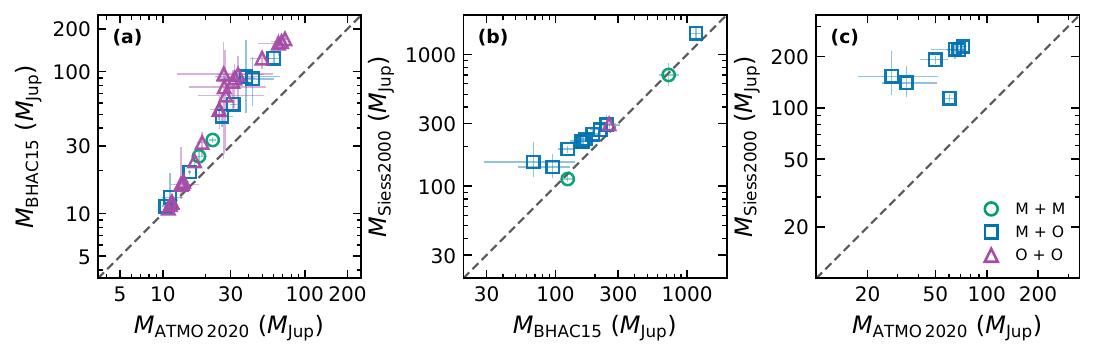}
\caption{
Comparison of the model-dependent masses obtained for the same sources using (a) ATMO~2020 and BHAC15, (b) BHAC15 and the Siess models, and (c) ATMO~2020 and the Siess models. Green circles denote sources whose masses are obtained by grid interpolation with both models, blue squares those using grid interpolation for one model and a temperature-based estimate for the other, and purple triangles those using temperature-based estimates for both models. The legend codes M and O denote matched and overluminous, respectively. For the blue squares, the horizontal-axis masses are temperature based, while the vertical-axis masses are obtained by grid interpolation. Sources with uncertain classifications are excluded. Error bars show the 16th--84th percentile intervals, and the dashed lines indicate equal masses.}
\label{fig:mass_comparison}
\end{figure*}

A source is classified as ``Matched'' when at least one model provides a valid \(T_{\rm eff}\)--\(L_{\rm bol}\) interpolation. Sources above the youngest isochrones without a valid interpolation are classified as ``Overluminous'', with masses estimated from the youngest \(T_{\rm eff}\)--mass relations. Sources lying below the \(10\) Myr model isochrones without a valid mass estimate are classified as likely ``Background'' stars because adopting the NGC~2024 distance would underestimate the luminosities of more distant field stars and place them below the young-model sequences. Sources whose median position lies outside the young-model regions but whose posterior distribution overlaps at least one model grid are labeled ``Uncertain''. They may still be NGC~2024 members because their positions can result from parameter uncertainties or limitations of evolutionary models at very young ages. The resulting sample contains {23} ``Matched'', {23} ``Overluminous'', {27} ``Background'', and {5} ``Uncertain'' sources. Table~\ref{tab:source_catalog} reports the source classifications, the mass-estimation method used for each evolutionary model, and the resulting model-dependent masses in \(M_{\rm Jup}\).

Five sources assigned to the ``Background” class (009, 057, 131, 227, and 492) have a thermal fraction of  \(f_{\rm th}^{1-2.6}\geq0.1\), with three reaching \(f_{\rm th}^{1-2.6}\geq0.3\). These sources are marked B* in Table~\ref{tab:source_catalog}. Strong near-infrared excess is commonly produced by warm dust in circumstellar disks around young stars \citep{2000AJ....120.1396H}. These sources may therefore be young NGC~2024 members with circumstellar material. The additional continuum can weaken photospheric features and bias the inferred parameters, making their classifications less reliable.

A recent study by \citet{Luhman2026} independently analyzed the same JWST/NIRSpec observations and assessed membership in NGC~2024 using near-infrared spectral features and other evidence of youth. Their catalog contains 87 sources, compared with 78 in this work, with 73 sources in common. Of the 14 sources appearing only in their catalog, ten were excluded during our initial screening because of incomplete spectral coverage. Among the remaining four, Sources 008 and 094 lack reliable photospheric features even after reprocessing by \citet{Luhman2026}. Source 251 is dominated by nebular emission. Source 666 is the only additional target with a substantially improved reprocessed spectrum and was not fitted separately in this work. Conversely, five sources appear only in our catalog, likely reflecting differences in spectral extraction and quality-screening criteria.

Of the 73 common sources, the classifications agree for 51 and differ for 22. The differences mainly arise from the membership criteria adopted by the two studies. We use spectral fits and evolutionary models, while \citet{Luhman2026} used visual comparisons with spectral standards and other evidence of youth. The largest group of discrepant sources contains ten objects classified here as NGC~2024 members but assigned undetermined membership by \citet{Luhman2026}. These sources are most highly extincted. Eight have \(A_{K,\rm small}+A_{K,\rm large}>3\) mag, while Sources 081 and 418 have approximately \(2.7\) and \(1.5\) mag, respectively. The extinction estimates reported by \citet{Luhman2026} are also generally high, suggesting that their near-infrared youth signatures may be obscured by such high extinction.

In the opposite direction, nine sources identified as members by \citet{Luhman2026} are classified here as either ``Uncertain'' ({067}, 137, 325, and 677) or ``Background'' ({009}, 069, 131, 140, and 492). The four ``Uncertain'' sources lie close to the young evolutionary grids, and their classifications mainly reflect model boundaries and parameter uncertainties. The five ``Background'' sources are likely members, but their fitted parameters may be affected by circumstellar or nebular emission. Sources {009}, 131, and 492 are already marked B$^{*}$ because of their {thermal contributions}. We further mark Sources 069 and 140 as B$^{*}$ because they have independent evidence of youth and residual nebular emission in their spectra. Their ``Background'' labels are retained because they are based on HR-diagram position. The remaining three discrepancies are Sources 023 and 320, classified here as ``Background'' but assigned undetermined membership by \citet{Luhman2026}, and Source 123, classified here as a member but not identified as such in their analysis. These discrepancies reflect the different methods adopted by the two studies, while the overall classifications remain broadly consistent.

After excluding the seven B$^{*}$ sources, the remaining {20} sources in the ``Background'' class all have \(f_{\rm th}^{1-2.6}<0.1\), limiting dilution of their photospheric features by thermal emission. Their inferred \(T_{\rm eff}\) and \(L_{\rm bol}\) are therefore less affected by thermal-continuum contamination. Despite the limited precision of individual fitted gravities, near-infrared gravity indicators can distinguish young low-gravity objects from higher-gravity field stars \citep{2003ApJ...593.1074G,2013ApJ...772...79A}. Figure~\ref{fig:logg_classification} shows that this cleaner Background sample has a median fitted \({\log g=4.34}\), compared with \({3.56}\) for the weakly veiled ``Matched'' and ``Overluminous'' candidate members. {The lower median gravity of the candidate members is consistent with their youth, whereas the higher median of the Background sample is compatible with older field dwarfs. The spread in fitted gravities may reflect both intrinsic differences among field stars and fitting uncertainties.}

\begin{figure*}[t]
\centering
\includegraphics[width=\textwidth]{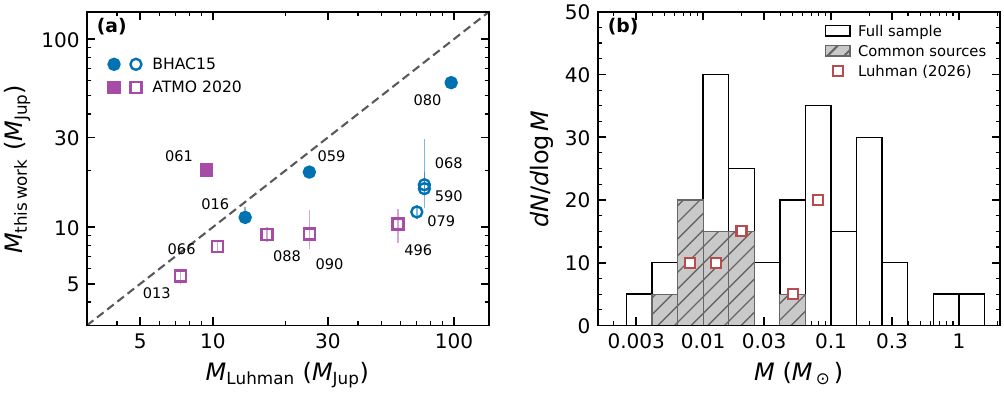}
\caption{
{Comparison of our mass estimates and mass distributions with those of \citet{Luhman2026}. (a) Mass estimates for 12 common sources. Blue circles and purple squares indicate masses derived using the BHAC15 and ATMO~2020 evolutionary models, respectively. Filled markers indicate joint \(T_{\rm eff}\)--\(L_{\rm bol}\) interpolation, while open markers indicate temperature-based estimates. Vertical error bars show our adopted mass uncertainties. The dashed line marks equal masses. (b) Mass distributions of the same 12 sources using our mass estimates (gray hatched histogram) and the estimates of \citet{Luhman2026} (red squares). The full sample of 46 candidate members in this work is shown as a white histogram with black outlines.}}
\label{fig:luhman_mass_comparison}
\end{figure*}

\subsection{Model Dependence of the Mass Estimates}
\label{subsec:mass_comparison}

Figure~\ref{fig:mass_comparison} compares the mass estimates for sources with valid estimates from more than one evolutionary model. After excluding uncertain cases, 31 sources have estimates from both ATMO~2020 and BHAC15, 13 from both BHAC15 and the Siess models, and seven from both ATMO~2020 and the Siess models. The ATMO~2020 and BHAC15 masses agree relatively well at the low-mass end but diverge toward higher masses. This mass-dependent trend is similar whether the sources are matched by both models, overluminous in one model, or overluminous in both models. The BHAC15 and Siess-model masses generally show closer agreement, whereas the ATMO~2020 masses are systematically lower than those inferred from the Siess models.

Model-dependent masses reflect the physical assumptions and parameter coverage of the adopted evolutionary calculations. They are also affected by the method used to estimate mass from each model grid. The first effect is most apparent in panel (a). ATMO~2020 is designed primarily for substellar objects and adopts an updated H--He equation of state, molecular opacities, and atmospheric boundary conditions \citep{2020A&A...637A..38P}. BHAC15 uses a different treatment of the interiors and atmospheric boundary conditions of brown dwarfs and low-mass stars \citep{2015A&A...577A..42B}. The two models agree relatively well at low masses but diverge toward the upper mass boundary of ATMO~2020.

The second effect is the mass-estimation method. Sources within a model grid use joint interpolation in \(T_{\rm eff}\) and \(L_{\rm bol}\), while those above the youngest isochrone use the youngest \(T_{\rm eff}\)--mass relation. In panel (b), BHAC15 and the Siess models agree more closely when both use grid interpolation. Their differences increase when BHAC15 provides a temperature-based mass and the Siess models provide an interpolated mass. For these sources, the temperature-based estimates have larger systematic uncertainties and are generally lower than the masses obtained by joint grid interpolation. The pronounced systematic offset in panel (c) reflects both the contrasting mass-estimation methods and the distinct physical assumptions of the two models. In particular, the ATMO~2020 grid extends only to approximately \(0.075\,M_\odot\), below the \(0.1\,M_\odot\) lower boundary of the Siess models. We therefore retain the mass estimates from each evolutionary model separately.

\subsection{{Reassessing the Mass Distribution of NGC~2024}}
\label{subsec:mass_distribution}

{We compare our mass estimates with those in Table~2 of \citet{Luhman2026} to examine differences in the inferred mass distribution. Some sources have mass estimates from multiple evolutionary models (Table~\ref{tab:source_catalog}). Among models providing valid joint \(T_{\rm eff}\)--\(L_{\rm bol}\) interpolation, we prioritize BHAC15, followed by ATMO~2020 and the Siess models. If no valid interpolation is available, we adopt temperature-based estimates using the same model priority.}

{\citet{Luhman2026} used evolutionary models to convert extinction-corrected absolute F444W magnitudes into masses, assuming an age of \(0.5\) Myr. Fourteen entries in their Table~2 appear in our fitted catalog. Sources 009 and 023 have no adopted young-member mass estimates in our analysis, leaving 12 sources for the comparison in the left panel of Figure~\ref{fig:luhman_mass_comparison}. The grid-interpolated estimates show closer agreement, whereas our temperature-based estimates are systematically lower. The latter apply to sources above the youngest isochrones in Figure~\ref{fig:teff_lbol}, whose elevated luminosities may reflect larger radii at very young ages \citep{2002A&A...382..563B}. Converting this enhanced brightness into mass at a fixed age can therefore overestimate their masses. Our prescription instead uses the youngest \(T_{\rm eff}\)--mass relation without translating the excess luminosity into a larger mass. Differences in the adopted evolutionary models and age treatment may also contribute to the offsets.}

{The right panel of Figure~\ref{fig:luhman_mass_comparison} compares the mass distributions of the common sources. Our distribution is shifted toward lower masses relative to that derived from the estimates of \citet{Luhman2026}, consistent with the source-by-source comparison. The mass distribution of our full candidate-member sample extends into the stellar-mass regime. At comparable apparent magnitudes, these highly extincted members have higher extinction-corrected luminosities, which, together with their fitted temperatures, yield larger inferred masses. By jointly constraining temperature, photospheric luminosity, and extinction, our approach also provides mass estimates for obscured candidates with uncertain spectral classifications or unavailable F444W photometry. Explicitly separating thermal excess from photospheric emission helps avoid attributing excess flux to the photosphere and overestimating luminosity-based masses. Additionally, our posterior-based analysis further propagates parameter uncertainties into the derived masses, providing individual uncertainty estimates.}

{The full-sample distribution shows an apparent peak near \(0.01\,M_\odot\), followed by a decline toward lower masses. This mass is close to the break reported by \citet{2025ApJ...981L..34D}. However, this similarity alone does not establish the presence of an intrinsic IMF turnover. In particular, a peak in our \(dN/d\log M\) distribution is not directly equivalent to a turnover in their \(dN/dM\) model. Our distribution also describes the observed sample rather than the intrinsic IMF, since effects such as photometric incompleteness, spectroscopic target selection, and extinction-dependent sensitivity have not been accounted for. More complete JWST spectroscopy of NGC~2024, designed with a well-defined selection function and combined with deep imaging, would provide a more representative sample. Applying this framework to such data would enable a more reliable determination of the cluster IMF.}

\section{Discussion}
\label{sec:discussion}

\subsection{Multi-component Modeling in Complex Clouds}
\label{subsec:complex_extinction}

The principal advantage of our multi-component framework is that it fits the atmospheric spectrum together with source-dependent dust and ice opacities. This flexibility is important in dense star-forming regions, where ice mantles, grain growth, and spatial variations in dust properties can produce extinction curves that are not captured by a fixed empirical law. Infrared extinction curves in dense molecular clouds, including NGC~2024/2023, are known to differ from those in the diffuse ISM \citep{2007ApJ...663.1069F,2009ApJ...690..496C}. If these variations are ignored, the fit may compensate for broad changes in the continuum by adjusting \(T_{\rm eff}\), extinction amplitude, or flux scaling, thereby biasing the inferred \(R\) and \(L_{\rm bol}\).

Our results directly show that the extinction curve varies among the NGC~2024 sightlines. {Eleven} sources are adequately reproduced with the baseline small-grain opacity, whereas {67} are better fitted with additional large-ice opacity; 11 of the latter require additional gray opacity, modeled here with large dust grains. As illustrated in Figure~\ref{fig:model_hierarchy}, these additional components reproduce broad changes in the H$_2$O ice-band profile and continuum curvature that cannot be removed by adjusting only the photospheric parameters or extinction amplitude. Within the multi-component framework, the model hierarchy allows the extinction shape to vary while {using BIC to balance likelihood improvement against the additional free parameters}. Allowing the extinction curve to vary reduces biases in the atmospheric parameters and flux scaling, leading to more reliable physical properties. 

Atmospheric-model systematics remain an additional source of uncertainty for the coolest sources. Incomplete treatment of photospheric dust and cloud opacity can be partly absorbed into an ISM-like reddening term, potentially biasing the inferred extinction magnitude \citep{2024ApJ...961..121H,2025ApJ...980..230T,2025ApJS..281...49T}. However, the reddening attributed to atmospheric-model limitations in these studies is small compared with the much larger extinction toward most of our sources, so this effect is unlikely to dominate the inferred extinction.

\subsection{Circumstellar Emission and Model Limitations}
\label{subsec:model_limitations}

\begin{figure}
\centering
\includegraphics[width=\columnwidth]{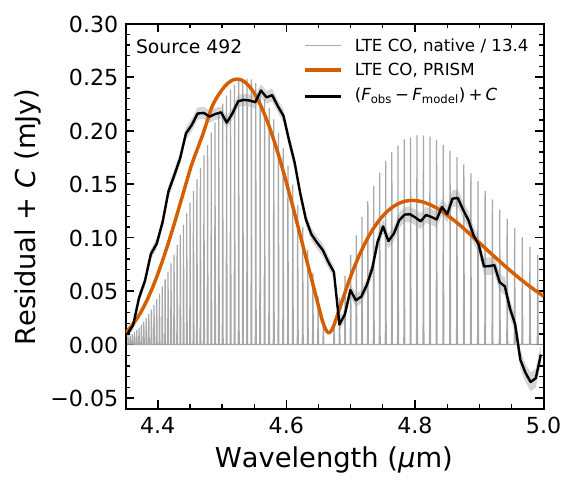}
\caption{
Residual spectrum (black) of Source 492 after subtracting the adopted model, with a constant offset applied to account for possible overestimation of the thermal continuum. The thin gray curve shows an illustrative \(T_{\rm CO}=1500\) K LTE CO \(v=1\)--0 spectrum constructed using CO line parameters from HITRAN2020 \citep{2022JQSRT.27707949G} and rescaled for display. The orange curve shows the same template degraded to the observed PRISM sampling. The residual morphology is broadly consistent with emission from warm CO gas.
}
\label{fig:co_gas_residual}
\end{figure}

The thermal component included for {39} sources accounts for smooth long-wavelength emission that is not reproduced by the photosphere and extinction alone. It represents continuum emission from warm circumstellar dust and adequately fits most sources requiring this component. However, YSO disks are generally gas rich, and molecular emission not included in the current model may contribute significantly to some spectra. The molecular gas in YSO disks is dominated by H$_2$, with CO as the principal observable carbon-bearing molecule; warm disk surface layers also contain H$_2$O and OH. Among these species, CO produces particularly strong near-infrared rovibrational emission from disks, envelopes, and outflows \citep{2011A&A...533A.112H,2015ApJ...809..167B}. 

{Sources~492 and~496 show clear CO-like residuals, as illustrated for Source~492 in Figure~\ref{fig:co_gas_residual}.} After subtracting the adopted model, the residual shows two broad peaks on either side of the gas-phase CO \(v=1\)--0 fundamental rovibrational band. We construct an illustrative \(T_{\rm CO}=1500\) K LTE spectrum using CO \(v=1\)--0 line parameters from the HITRAN2020 database \citep{2022JQSRT.27707949G}. The native spectrum contains numerous narrow rovibrational lines distributed between the R- and P-branches. After convolution to the observed NIRSpec/PRISM sampling, these lines blend into two broad structures that qualitatively resemble the overall morphology of the observed residual. This similarity suggests that unresolved warm-CO emission contributes to the spectrum. Because the forward model does not include molecular gas emission, part of the unresolved CO emission may be incorporated into the fitted thermal continuum. The thermal contribution may therefore be overestimated, producing the mismatch toward \(5~\mu{\rm m}\) after model subtraction. Additional differences may arise because the illustrative template assumes a single-temperature LTE gas and does not include optical-depth or kinematic effects.

We do not include an explicit molecular-gas component in the present model because coherent CO-like residuals are detected in only a small subset of the spectra. {The residuals are concentrated mainly in the \(4.4\)--\(5.0~\mu{\rm m}\) region}, whereas \(T_{\rm eff}\) is constrained primarily by the short-wavelength photospheric features and extinction by the overall continuum shape. Unmodeled CO emission is therefore unlikely to substantially alter these parameters, but it can directly affect the normalization and inferred properties of the thermal component, which is only phenomenological and is not a primary focus of this work. For sources with very strong molecular emission, an explicit gas-emission component would be required to obtain a better spectral fit.

\section{Summary and Conclusions}
\label{sec:conclusions}

We present a catalog of spectroscopically inferred properties for sources observed toward NGC~2024 with JWST/NIRSpec. This catalog is derived from multi-component forward modeling of atmospheric spectra, dust and ice opacity, and optional thermal emission. Of the 137 extracted spectra, 85 pass the initial quality screening and 78 have adopted fits. {Eight} spectra with \({\rm MAF}>0.05\) are identified by fit-quality flags. Our principal results are as follows.

\begin{enumerate}

\item Most spectra require extinction curves more complex than the baseline small-grain model. Additional large-H$_2$O-ice opacity is favored for {67} sources, while {11} are adequately reproduced by the baseline model. Eleven sources require additional gray opacity, which is modeled with large refractory grains. A thermal component is included for {39} sources.

\item The catalog reports posterior constraints on the atmospheric parameters, extinction, grain properties, and photospheric flux scaling. Combining these parameters with the distance to NGC~2024 yields radii and bolometric luminosities. Model assignments, derived parameters, and quality flags are provided in machine-readable form.

\item Comparison with evolutionary models for young low-mass and substellar objects identifies 46 candidate members, {27} sources classified as ``Background'', and {5} uncertain sources. Seven of the Background sources are marked B$^{*}$ because independent evidence suggests membership. Of the candidate members, {23} are matched by at least one model grid and {23} lie above the youngest available isochrones. The latter may reflect the young age of NGC~2024 and uncertainties in model initial conditions. Candidate members also show lower fitted surface gravities than the remaining Background sources.

\item We estimate masses of candidate members using joint temperature--luminosity interpolation, adopting the youngest temperature--mass relations for sources above the youngest isochrones. Modeling the extinction toward each source extends our characterization to highly obscured candidates in the stellar-mass regime. Separating thermal excess helps avoid luminosity-based mass overestimates, while temperature-based estimates for overluminous sources yield lower masses than fixed-age photometric estimates. Masses from the three evolutionary models are reported separately.

\item The phenomenological thermal component reproduces most smooth long-wavelength excesses. A small subset shows residual structure resembling unresolved warm-CO emission. This unmodeled emission is expected to affect the inferred thermal component more directly than the atmospheric and extinction parameters.

\end{enumerate}

Our multi-component framework jointly models atmospheric spectra, dust opacity, ice opacity, and circumstellar emission. Applied to NGC~2024, it yields a uniform set of atmospheric parameters and derived physical properties while allowing extinction and circumstellar emission to vary among sources. {Compared with approaches based on spectral standards and an adopted extinction law, this decomposition helps characterize highly extincted candidates and distinguish thermal excess from photospheric emission. The common-source comparison with \citet{Luhman2026} further illustrates how the adopted spectral analysis and mass-estimation method can affect individual masses and the inferred mass distribution.} The resulting catalog also provides a basis for studying ice and dust-grain properties in NGC~2024. The same approach can be applied to JWST spectroscopic surveys of other obscured star-forming regions.

\begin{acknowledgments}

{We thank the referee for helpful comments and suggestions.} This work is supported by the National Natural Science Foundation of China (NSFC) through projects 12622304, 12633007, 12373028, 12588202, 12603059, 12322306, and 12133002. S.W. and X.C. acknowledge support from the Youth Innovation Promotion Association of the Chinese Academy of Sciences under grant Nos. 2023065 and 2022055, respectively. Z.T. acknowledges support from the Postdoctoral Fellowship Program of CPSF under Grant Number GZC20261725. This work is based on observations made with the NASA/ESA/CSA James Webb Space Telescope. The data were obtained from the Mikulski Archive for Space Telescopes at the Space Telescope Science Institute, which is operated by the Association of Universities for Research in Astronomy, Inc., under NASA contract NAS 5-03127 for JWST. These observations are associated with program 5409 {(\dataset[doi:10.17909/n4d1-sn72]{https://doi.org/10.17909/n4d1-sn72})}.

\end{acknowledgments}

\facilities{JWST (NIRSpec)}
\bibliography{NGC2024_parameters}{}
\bibliographystyle{aasjournalv7}



\end{document}